\documentclass[11pt,html,nofootinbib,preprint,prd,aps]{revtex4} 
\usepackage{graphicx}
\input html.sty
\begin{document}
\title{Free will, undecidability, and the problem of time in quantum gravity}

\author{Rodolfo Gambini}
\address{Instituto de F\'{\i}sica, Facultad de Ciencias, 
Universidad
de la Rep\'ublica, Igu\'a 4225, CP 11400 Montevideo, Uruguay\\
rgambini@fisica.edu.uy}
\author{Jorge Pullin}
\address{Department of Physics and Astronomy, 
Louisiana State University, Baton Rouge,
LA 70803-4001\\pullin@lsu.edu}

\date{September 5th 2008}

\begin{abstract}
In quantum gravity there is no notion of absolute time. Like all other
quantities in the theory, the notion of time has to be introduced
``relationally'', by studying the behavior of some physical quantities
in terms of others chosen as a ``clock''. We have recently introduced
a consistent way of defining time relationally in general
relativity. When quantum mechanics is formulated in terms of this new
notion of time the resolution of the {\em measurement problem} can be
implemented via decoherence without the usual pitfalls. The resulting
theory has the same experimental results of ordinary quantum
mechanics, but every time an event is produced or a measurement
happens two alternatives are possible: a) the state collapses; b) the
system evolves without changing the state. One therefore has two
possible behaviors of the quantum mechanical system and physical
observations cannot decide between them, not just as a matter of
experimental limitations but as an issue of principle.  This
first-ever example of fundamental undecidability in physics suggests
that nature may behave sometimes as described by one alternative and
sometimes as described by another.  This in particular may give new
vistas on the issue of free will.
\end{abstract}

\maketitle
\eject

In general relativity there is no notion of absolute time. In fact,
there is no absolute notion. All physical predictions have to be
formulated as relations between physical quantities.  This has been
recognized since the early days of general relativity through
Einstein's ``hole argument'' \cite{hole}. In particular the notion of
time has to emerge ``relationally''. One possible way of attacking this
was introduced by Page and Wootters \cite{PaWo}.  In their
proposal one takes any physical quantity one is interested in studying
and chooses another physical variable that will act as ``clock''. One
then studies how the first variable ``evolves as a function of the
second one''. In this view, time does not play any preferred role among
other physical variables. This is in contrast to 
ordinary quantum mechanics where
one has to unnaturally assume that time is supposed to be the only
variable in the universe not subject to quantum fluctuations. In spite
of the simplicity and naturalness of this proposal to tackle the
problem of time in quantum gravity, technical problems arise. The
problems are related with what one considers to be physical quantities
in a theory like general relativity. Usual things that one may
consider physical quantities, like ``the scalar curvature of
space-time at a given point'' are not well defined objects in general
relativity.  The problem is what is ``a given point''? Points in space
have to be defined physically in general relativity. One can
characterize a point as a ``place where something physical happens''
(for instance, a set of physical fields takes certain values). Then
one could ask ``how much is the curvature at that point''. The end
result is indeed physical. But it is again a relation between the
values of curvatures and fields. Such relation is given and
immutable. How could one construct a clock out of something immutable?
It appears that the only things that are physical are immutable
relations and the only things that evolve are the members of the
relations, like the curvatures and fields. In technical terms, what
one can consider as physical observable in general relativity is a
quantity that is left invariant under the symmetries of the theory, or
in the canonical language, that commutes with the constraints. Since
one of the constraints is the Hamiltonian, physical quantities do not
evolve. Therefore they cannot work as clocks. This created problems
\cite{kuchar}
for the Page--Wootters proposal.  A way out was sought by trying to
establish relations not between physical quantities but between
mathematical quantities one uses to describe the theory that are not
directly measurable (like for instance, the components of the metric
at a point). Far from helping, this led to significant technical
problems since one ends formulating the theory in terms of
unobservable quantities. Ultimately it was shown in model systems that
the proposal cannot be used to compute elementary things, for instance
quantum probabilities of transition \cite{kuchar}.

The observation we have recently made \cite{time} is that the
Page--Wootters construction can be rescued by using Rovelli's proposal
of ``evolving constants of the motion'' \cite{rovelli}, a concept that
can be traced back to DeWitt, Bergmann and Einstein himself.  This
idea is to introduce genuinely observable physical quantities,
i.e. relations between magnitudes as we highlighted above, but that
depend on a continuous parameter. If one imagines evolution as changes
in such a parameter, one can actually construct the relational
description of Page and Wootters and show that it actually leads to
the correct quantum probabilities of transition, at least in model
systems \cite{time}. The beauty of the complete construction is that
the continuous parameter in the evolving constants completely drops
out at the end of the day and the formulation remains entirely written
in terms of truly observable physical quantities, even in the extreme
situations that can develop in physics when quantum gravity effects
become important.

Remarkably, taking seriously this relational solution  to the problem of time in quantum gravity
has implications for all of physics. We will concentrate on the
the changes that this new description introduces in ordinary quantum
mechanics. When one has a relational description where one's ``clock''
is a physical variable subject to quantum fluctuations like any
any other, the description of ordinary quantum mechanics is
different from the traditional one. In the ordinary formulation
there exists an absolute time that is {\em not
quantum mechanical}, it is represented by a completely classical
parameter that is not subject to quantum fluctuations.  This is
clearly an idealization since all clocks we use in the real world are
subject to such fluctuations. Although very small to be observed in
practice, they are there.

At this point it is good to introduce another element that will be
modified by the use of real clocks and rods to describe quantum
mechanics\footnote{We concentrate in this essay on the use of real
clocks, but similar effects appear in the measurement of distances
that add on to the ones we discuss here. See \cite{tangle} for
details.}.  The evolution of states in ordinary quantum mechanics is
technically called ``unitary''. This in particular implies that
information is preserved in evolution. In a sense ``evolution is
trivial'' and everything is determined given the initial state. The
only place where something non-trivial happens in ordinary quantum
mechanics is when a measurement takes place. There the quantum states
are supposed to evolve in a non-trivial way. Why this type of
evolution happens in the measurement process and what justifies such
change in the state is one of the open conceptual problems of quantum
mechanics called {\em the measurement problem} \cite{measurementwiki}.

Acknowledging that the real clocks and rods that one may use to
measure space-time are not arbitrarily accurate requires reformulating
the theory in terms of such clocks and rods. We have carried out such
reformulation in some detail in ref. \cite{obregon}. It is not too
surprising that in the resulting picture one does not have a unitary
evolution: although the underlying theory {\em is unitary}, our clocks
and rods are not accurate enough to give a depiction of evolution that
{\em appears unitary}. It is also no more the case that evolution is
only non-trivial at the measurement of quantities. This also implies
that there is a steady loss of information in the evolution of quantum
states.

The reader at this point may ask: sure, a classical non-fluctuating
time is an idealization. But one uses idealizations frequently in
physics. Can I not always find a clock such that its quantum
fluctuations are small enough to ignore this effect altogether? The
answer is negative. And it is quantum mechanical and gravitational in
nature.  It is well known that in quantum mechanics one needs to
expend energy in order to achieve accurate measurements
\cite{saleckerwigner}. On the other hand, gravity puts fundamental
limits on how much energy can be concentrated in a measuring device
before it turns into a black hole.  Coupling together these two
observations, one concludes that there exist fundamental limits,
imposed by quantum mechanics and gravity, on how accurately we can
measure distances and time \cite{karolyhazy,ng}.  A detailed
calculation shows that these limits imply that quantum states
described in terms of a realistic clock variable $T$ lose coherence
(quantum information) exponentially. The exponent is given by $T_{\rm
  Planck}^{4/3} T^{2/3} \omega^2$ where $T$ is the time elapsed
$\omega$ is the Bohr frequency associated with the energy difference
between components of the quantum state (to have evolution one needs a
superposition of states with different energies) and $T_{\rm
  Planck}\sim 10^{-44}s$ is Planck's time\footnote{Although the
  particular time dependence has been questioned \cite{requardt}, for
  what follows one only needs an effect that grows with time and other
  mechanisms have been proposed that yield such behavior, albeit with
  a weaker time dependence \cite{kudaka}.}. The effect is too small to
be observed with current technologies, but might be within the reach
of technologies of the relatively near future \cite{SiJa}. To give an
idea of the meaning of these numbers, the loss of coherence is larger
the larger the energy difference of the quantum states in
superposition is and the longer one waits. To have something visible
in typical times in the lab, one would require states involving about
$10^{12}-10^{14}$ atoms in coherence. Current Bose--Einstein
condensates have $10^6$ atoms in coherence. Notice that although the
effect is too small to be observed today, its existence is not
controversial, since one can magnify the behavior arbitrarily just by
choosing bad clocks. In fact, experiments in cavities can be
reinterpreted in this way and the effect is readily measurable
\cite{bonifacio}.

Although experimentally not detectable today, this fundamental effect
has important conceptual implications, in particular for the {\em
measurement problem in quantum mechanics} we mentioned above.  As
stated, the latter refers to the fact that the state of the system
being measured changes abruptly during the process of measurement.
Technically it falls into an {\em eigenstate of the measured quantity}
right after a measurement has been performed. This is usually referred
to as the ``reduction process''. The conceptual problem is: how can
one explain this abrupt change of state?  Notice that this problem is
quite pervasive because in quantum mechanics ``measurement'' has a
more general meaning than in common parlance.  Namely, in quantum
mechanics the theory describes probabilities of events. Every time an
event happens, a ``measurement'' takes place. For instance, every
element of the reality we see around us is constituted by a network of
``events'' and therefore may be considered as a result of many
``measurements'' (Tegmark has a nice discussion of this point
\cite{tegmark}).

A widely accepted explanation of the abrupt change in the wavefunction
in a measurement is to consider that there exists an interaction
between the system being measured and the environment in general (in
particular with the measuring device). Both the environment and the
measuring device are typically systems with a vastly larger number of
degrees of freedom than those of the quantum system being studied. 
Also, the measuring device has many more degrees of freedom than
the one displaying the measurement (e.g. ``the needle in a gauge'').
The interaction of the quantum system with the environment and
measuring device leads to the quantum system losing information that
in particular is not registered in ``the needle of the gauge''.  The
end result is that the evolution apparently loses unitarity, and
information appears to be lost. Because the interaction is with a vast
number of degrees of freedom, the rate of information loss is very
quick and the phenomenon therefore appears as ``abrupt''.

The above ``solution'' to the measurement problem has been criticized.
We cannot do justice to the full extent of the problem and its
associated vast literature in the confines of this essay (see for
instance the book of d'Espagnat \cite{despagnat} for references).  Our
claim is that the fact that non-trivial evolution happens all the time
due to our imperfect clocks and measuring rods contributes to
surmounting a significant portion of these obstacles \cite{measurement}. 
Let us briefly review how.

There are two main criticisms to the solution of the problem of
measurement by invoking the environment (decoherence). The first
criticism is that that the ``system plus environment'' evolves
unitarily and therefore all information is still present at the end
and could in principle be retrieved. The second type of criticism is
related to the fact that the system is left in a superposition of
states through the interaction with the environment and therefore it
would not generate a definite event (or measurement) but a
superposition of them.

Let us address the first of the two criticisms.  No matter how large
the number of degrees of freedom of the environment, one can {\em in
principle} recover the information by realizing a measurement of the
joint ``system plus measuring device plus environment'' system
\cite{despagnat}. Another way of recovering the information, assuming
the ``system plus measuring device plus environment'' is a closed
system is  to just wait for a long time. Eventually the interactions of
the system with the environment and measuring device will bring back
the information to the ``system plus measuring device'' degrees of
freedom. These phenomena are called ``revivals'' in the literature
\cite{despagnat}. 
But let us now reconsider the fundamental loss of coherence that
arises due to our inability to measure space-time with arbitrary
precision. It has the attractive feature of killing off the
possibility of ``revivals'' in a fundamental, inescapable way. If one
attempted to ``wait longer'' to see revivals, the effect we discussed
just becomes larger, as we pointed above.  Therefore the detailed study of
concrete examples like Zurek's model (see \cite{measurement} for
details) lead us to conclude that the longer one waits the more
information the system loses and the chances of revivals actually
diminish.  One can also see that the effect does not allow to measure
observables for the complete system including the environment and
attempting to recover the information that way. The reason for this is
that the fundamental loss of coherence that we discuss in this essay,
although typically minute, is magnified 
---in examples we have studied--- due to the large number of
degrees of freedom that get involved in a measurement process
\cite{measurement2}. 

Before addressing the second objection, let us note that the above
behavior naturally leads to the main point of the essay:
undecidability.  Since examples have led us to the conjecture that one
cannot measure observables of the whole system plus environment nor we
can observe revivals one cannot decide if the quantum state has
suffered reduction or it evolved unitarily. In fact, it could even be
conceivable that sometimes there might be reduction, sometimes not,
and we do not have reasons to expect one or the other in a given
instance.  The difference between these two views of nature can be
very significant. In one extreme case quantum states are given ``once
and for all'' as initial conditions. The evolution is unitary, and
what we perceive as loss of unitarity is due to our inability to
access the underlying variables of the theory, due to gravitational
limitations.  In the other extreme view, quantum states are evolving
all the time due to reduction processes.  There can also be
combinations of the two scenarios where in some events evolution is
unitary and in others is not.  Undecidability does not imply that the
difference between these two scenarios is irrelevant. For instance,
there may exist complex systems (an intriguing example are living
organisms) for which it is impossible to prepare their initial state
or consider ensembles.  For such systems, undecidability may occur
among widely different states.  No matter what is the outcome,
reduction or unitary evolution, the choice between these alternatives
could produce observable phenomena later on. The specific outcome may
have important consequences on the occurrence of future events.

Let us get back to the second objection to the decoherence solution to
the measurement problem: that at the end of the interaction with the
environment, the measuring apparatus is generically left in a
superposition of (eigen)-states corresponding to different ``positions of
the needle of the gauge''. That would not correspond to what one usually
calls a ``measurement'' in which the apparatus is in a given 
(eigen-)state,
corresponding to the ``needle of the gauge'' taking a definite
position. This is what John Bell called the ``and-or'' problem
\cite{bell}. Namely, What explains that the needle took a definite
position from within the superposition of states?  In particular, did
a further change in the state occur to select the given position of
the needle?  This does not have to be the case. For instance, in the
many-worlds interpretation of quantum mechanics it is assumed that the
state has a unitary evolution all the time but it does not describe a
particular universe but the whole set of alternatives. In our world
only one of the states in the superposition is realized and in each
world a different event is observed. Also, in the modal
interpretations \cite{modal} the occurrence of events is associated to
the ``actual properties'' that the system can acquire without changing
its state, which evolves unitarily.  It turns out that our effect may
help avoid this problem since it allows to define the appearance of
events without necessarily implying a change in the quantum state. We
can assume that an event occurs when the distinction between the
``system plus apparatus plus environment'' being in a superposition or
in a given state becomes undecidable. This phenomenon is typical of
interaction with an environment or a measuring device, i.e. it does
not occur in quantum systems in isolation. In such case a unitary
evolution or an abrupt change as the one given by a collapse would be
obviously distinguishable.

The above proposal leads naturally to a revision of the ideas of
natural laws in physics. In philosophy there are different attitudes
that have been taken towards the physical laws of nature (see for
instance
\cite{stanford}).  One of them is the ``regularity theory'', many times 
attributed to Hume \cite{hume}; in it, the laws of physics are
statements about uniformities or regularities of the world and
therefore are just ``convenient descriptions'' of the world. Ernest
Nagel in {\em The Structure of Science} \cite{nagel} describes this
position in the following terms:  {\em ``Hume proposed an analysis
of causal statements in terms of constant conjunctions and de facto
uniformities.. ---according to Hume [physical laws consist] in certain
habits of expectation that have been developed as a consequence of the
uniform but de facto conjunctions of [properties]."}  The laws of physics
are dictated by our experience of a preexisting world and are a
representation of our ability to describe the world but they do not
exhaust the content of the physical world.

A second point of view sometimes taken is the ``necessitarian theory''
\cite{stanford}, which states that laws of nature are ``principles''
which govern the natural phenomena, that is, the world ``necessarily
obeys'' the laws of nature. The laws are the cornerstone of the
physical world and nothing exists without a law. The presence of the
undecidability we point out suggests strongly that the ``regularity
theory'' point of view is more satisfactory since the laws do not
dictate entirely the behavior of nature.

Let us turn now our attention to the issue of freedom.  We have seen
that after the occurrence of an event the system may choose between
behaving as if there has been a reduction process or not.  That is,
after the observation of the event either the system simply behaves as
if it were part of the universe and its state were that of the
universe or if as its state would be given by the reduction
postulate. In the first case the system would keep its entanglement
with the rest of the universe (i.e. the environment), in the second it
will lose its entanglement. The availability of this choice opens the
possibility of the existence of free acts. 
This type of act of the system will not imply any violation whatsoever of the
laws of physics, understanding the latter as regularities in the
observation of nature. It should be noted that this freedom in the
system is not even ruled by a law of probabilities for the possible
outcomes.

It is worthwhile pointing out that the notion of free will introduced
here is different from the one
introduced by Conway and Kochen (CK)
\cite{freewill}. We have started from quantum mechanics and 
gravitation and concluded that there exists undecidability and as a
consequence free will. CK, on the other hand, start by considering a
human observer conducting an experiment, which can make one of a set
of possible observations. Starting from this and a limited set of
assumptions that do not involve assuming that quantum mechanics holds,
they conclude that elementary particles and other microscopic systems
must also behave with ``free will''. 
This observation is 
attractive because it has as almost inevitable result that physics is
indeterministic in the sense that is usually understood in quantum
mechanics.

Freedom affects the causal structure of the world and therefore it
does not belong in the realm of psychology but the ultimate discussion
about its existence belongs in the realm of physics. The great
philosopher Spinoza was the first in successfully building a complete
philosophical system consistent with the laws of physics of his time.
Those laws were completely deterministic.  In his point of view, {\em
  ``in nature there is nothing contingent, but all things have been
  determined from the necessity of the divine nature to exit and
  produce an effect"} \cite{spinoza}.

We now have more advanced physical laws than the ones available to
Spinoza, and they seem to imply undecidability and allow for free
will.  
We live in a contingent world.  In it, the
transition from ``what could be'' to ``what is'' results either from mere
chance or from a meaningful choice of free will. It is surprising that
the freedom stemming from the undecidability yields two alternatives,
the choice between which is meaningful: either the systems involved in
events conserve their entanglement with the universe
or break that entanglement.
We would like to put forward the proposal that adopting the regularist
point of view together with the idea of undecidability may allow to
confront important objections to the libertarian\footnote{
Libertarianism in this context is a philosophical position that states
that human beings have free will and that the latter is incompatible
with determinism. This is usually interpreted to imply that
determinism is false.  Among the exponents of this point of view
(``incompatibilists'') are Peter van Inwagen, Robert Kane, Laura Ekstrom,
Timothy O'Connor and Thomas Pink
\cite{incompatibilist}.} stance. These types of objections have been
repeatedly leveled against attempts to substantiate free will based on
the probabilistic nature of quantum mechanics. In fact if quantum
mechanics only implies a mere lack of causal determination in the
occurrence of events, this is not sufficient to ensure that it makes
sense to consider a free act for which responsibility is possible.
These objections stem from the potential fallacy of considering that
only two exclusive alternatives exist: the deterministic and the
random, excluding the possibility that the agent have any capability
to control or self-determination over her acts.  Implicit in this
argument is the necessitarian point of view which excludes all aspects
of reality not controlled by physical laws.

To conclude, we have observed that inherent limitations in the
measurement process introduced by the use of a relational notion of
time in quantum gravity appear to imply undecidability in the laws of
physics.  This strengthens the regularist vision of physical laws and
opens the door to an essential difference through which free acts lead
to different possible evolutions of the quantum state when an event
takes place.  The ability to act freely we discuss stems from quantum
mechanics and therefore has a universal character. It is not entirely
clear that it is connected with the decision making process of humans.
It is currently widely contentious if quantum mechanics plays any role
in processes in the human brain \cite{tegmark2}. It would be quite
disappointing if a universe that naturally includes in the laws of
physics the capability for free acts will end up disallowing them for
human beings.

This work was supported in part by grants NSF-PHY0650715, and by funds
of the Horace C. Hearne Jr. Institute for Theoretical Physics, FQXi,
PEDECIBA and PDT (Uruguay) and CCT-LSU.

\end{document}